\documentclass[12pt]{article}
\usepackage{a4wide}
\usepackage{amssymb}
\begin{document}
{\renewcommand{\thefootnote}{\fnsymbol{footnote}}
\hfill  IGPG--07/6--5\\
\medskip
\begin{center}
{\LARGE Cosmological vector modes and quantum gravity effects}\\
\vspace{1.5em}
Martin Bojowald\footnote{e-mail address: {\tt bojowald@gravity.psu.edu}}
and 
Golam Mortuza Hossain\footnote{e-mail address: {\tt hossain@gravity.psu.edu}} \\
\vspace{0.5em}
Institute for Gravitation and the Cosmos,\\
The Pennsylvania State
University,\\
104 Davey Lab, University Park, PA 16802, USA\\
\vspace{1.5em}
\end{center}
}

\setcounter{footnote}{0}

\newtheorem{theo}{Theorem}
\newtheorem{lemma}{Lemma}
\newtheorem{defi}{Definition}

\newcommand{\proofend}{\raisebox{1.3mm}{\fbox{\begin{minipage}[b][0cm][b]{0cm}
\end{minipage}}}}
\newenvironment{proof}{\noindent{\it Proof:} }{\mbox{}\hfill \proofend\\\mbox{}}
\newenvironment{ex}{\noindent{\it Example:} }{\medskip}
\newenvironment{rem}{\noindent{\it Remark:} }{\medskip}

\newcommand{\case}[2]{{\textstyle \frac{#1}{#2}}}
\newcommand{\lP}{\ell_{\mathrm P}}

\newcommand{\md}{{\mathrm{d}}}
\newcommand{\tr}{\mathop{\mathrm{tr}}}
\newcommand{\sgn}{\mathop{\mathrm{sgn}}}

\newcommand*{\R}{{\mathbb R}}
\newcommand*{\N}{{\mathbb N}}
\newcommand*{\Z}{{\mathbb Z}}
\newcommand*{\Q}{{\mathbb Q}}
\newcommand*{\C}{{\mathbb C}}

\begin{abstract}
  In contrast to scalar and tensor modes, vector modes of linear
  perturbations around an expanding Friedmann--Robertson--Walker
  universe decay. This makes them largely irrelevant for late time
  cosmology, assuming that all modes started out at a similar
  magnitude at some early stage. By now, however, bouncing models are
  frequently considered which exhibit a collapsing phase. Before this
  phase reaches a minimum size and re-expands, vector modes grow. Such
  modes are thus relevant for the bounce and may even signal the
  breakdown of perturbation theory if the growth is too strong. Here,
  a gauge invariant formulation of vector mode perturbations in
  Hamiltonian cosmology is presented.  This lays out a framework for
  studying possible canonical quantum gravity effects, such as those
  of loop quantum gravity, at an effective level. As an explicit
  example, typical quantum corrections, namely those coming from
  inverse densitized triad components and holonomies, are shown to
  increase the growth rate of vector perturbations in the contracting
  phase, but only slightly. Effects at the bounce of the background
  geometry can, however, be much stronger.
\end{abstract}

\section{Introduction}

Linear perturbations on isotropic expanding or contracting geometries
can be split in different types according to their transformation
properties under spatial rotations: scalar, vector and tensor modes.
In general, this presents a convenient decomposition of general
perturbations into different classes. But it acquires a more important
role for the linearized Einstein's equation where the different modes
decouple. Moreover, linearized gauge transformations corresponding to
changes of space-time coordinates do not mix the modes which can thus
be analyzed separately. In this article we focus on the vector mode
and possible quantum effects in its evolution.

Vector mode perturbations in classical cosmology decay in an expanding
universe, and dynamically become of less interest as the universe
continues its expansion. Accordingly, vector modes are often ignored.
This feature holds true for expanding cosmologies which start from a
big bang or emerge from a quantum state if one assumes all modes to be
of comparable initial magnitude.  However, this assumption has to be
justified. One possibility is to use bouncing models where the
pre-history before the big bang is described by a collapsing phase.
Then, vector perturbations can lead to significant problems because
gauge invariant measures of vector perturbations grow. Their current
size relative to that of scalar and tensor modes then depends on where
equal sizes or other initial conditions are assumed. Moreover, vector
modes are generated by higher order perturbations and subsequently
grow in a contracting phase \cite{NonlinVector}. They can thus not be
ignored altogether.

There are several models such as string inspired pre-big bang
scenarios \cite{Veneziano,VenezianoLesHouches} or cyclic and ekpyrotic
models \cite{Ekpyrotic,CyclicEkpy}, or some models of loop quantum
cosmology \cite{Oscill,APSCurved} which exhibit a bounce. Such
scenarios for an avoidance of the big bang singularity are developed
mainly based on homogeneous models of cosmology.  On the other hand,
the growth of vector perturbations in the contracting phase indicates
a possible violation of the homogeneity assumption when the bounce is
approached as indicated by the breakdown of classical perturbation
theory. Thus, the growth of vector perturbations not only raises
questions regarding the validity of an homogeneity assumption but may
even question the phenomena of the bounce itself. \footnote{Based on
   matching conditions, an evolution of vector modes through a bounce
   has been studied, e.g., in \cite{VectorBounce,Giovannini:Vector},
   assuming certain forms of a non-singular bouncing background
   geometry. This does not address the validity of a perturbativity
   assumption. Vector modes have occasionally been considered in
   inflationary scenarios such as in \cite{HHR}. Since this happens in an
   expanding phase, vector modes decay and do not challenge the
   perturbativity assumption either.}

In this paper, we derive the vector mode dynamics in the context of
cosmological models based on loop quantum gravity
\cite{Rov,ALRev,ThomasRev} and cosmology \cite{LivRev}. Firstly, we
present a systematic derivation of classical vector perturbation
equations using a canonical formulation in Ashtekar variables
\cite{AshVar,AshVarReell}. We compute the gauge transformations
property for the vector perturbations and then construct the
corresponding gauge invariant variable. All this is done in a purely
canonical way to outline the general procedure followed also in the
presence of quantum corrections.

In the following section we study possible effects of quantum
corrections expected from loop quantum gravity. We compute
requirements on quantum correction functions from anomaly cancellation
in the quantum corrected constraint algebra. 
These expressions include inhomogeneities in such a way that all
symmetries of a Friedmann--Robertson--Walker background are broken.
Thus, we are not dealing with a mini-superspace quantization even
though inhomogeneities are restricted to the perturbative vector mode.
How individual terms of effective constraints are related to operators
in the full theory is described in \cite{QuantCorrPert}.
We study the effects of quantum correction arising for inverse
densitized triads in detail, and perform corresponding calculations
for corrections from holonomies in a subsequent section. While current
methods are not sufficient to compute full correction functions for
all gauges, it turns out that the remaining freedom is constrained by
requiring the absence of anomalies as a consistency condition. This
provides evidence, at the effective level employed here, that anomaly
cancellation will restrict possible quantization choices of the full
theory. (See \cite{AlexAmbig,AmbigConstr} for analogous statements
based on different principles.) Moreover, we demonstrate explicitly
that anomaly-free set of constraints, and thus a covariant
effective space-time picture, is possible even in the presence of
non-trivial quantum corrections.  As one application, we show that in
perturbative regimes (not close to a bounce) quantum corrections make
the growth rate of the vector mode in a contracting universe slightly
stronger compared to the classical behavior.

\section{Canonical formulation}

In this paper, we study the vector mode of linear metric perturbations
around spatially flat Friedmann-Robertson-Walker (FRW) spacetimes. 
The procedure we use is analogous to that for scalar modes
\cite{HamPerturb}, although specifics certainly change when vector
modes are considered.
The general form of the perturbed metric around the FRW background
containing only the vector mode is given by
\begin{equation} \label{PMVector}
g_{00} = -a^2 \quad,\quad g_{0a} = a^2 S_a \quad,\quad
g_{ab} = a^2 \left[\delta_{ab} + F_{a,b} + F_{b,a} \right] ~. 
\end{equation}
The perturbation fields $F^a$ and $S^a$ satisfy ${F^a},_{a} = 0$ and
${S^a},_{a} = 0$ to separate them from scalar gradient terms. In other
words these divergence free fields describe the vorticity of metric
perturbations. An index $0$ refers to conformal time $\eta$, while
$a,b,\ldots$ refer to co-moving spatial coordinates.  In a canonical
formulation, the distance element is expressed in terms of the spatial
metric $q_{ab}$, the lapse function $N$ and the shift vector $N^a$,
related to the spacetime metric through
\begin{equation} \label{MetricRelation}
g_{00} = -N^2+q_{ab}N^aN^b \quad,\quad g_{0a} = q_{ab}N^b
\quad,\quad g_{ab} = q_{ab} ~.
\end{equation}
By comparing expression (\ref{PMVector}) for a perturbed spacetime
metric to the relation (\ref{MetricRelation}), one can see that in
canonical formulations the vector perturbation is generated through
the perturbations of the shift vector $N^a$ and off-diagonal
components of the spatial metric $q_{ab}$. In particular, the lapse
function, being scalar, does not contribute to vector mode dynamics.

\subsection{Background} 

In canonical gravity, the spatial metric $q_{ab}$ plays the role of a
configuration variable with momenta related to extrinsic curvature
\begin{equation} \label{Kab}
 K_{ab}= \frac{1}{2N}\left(\dot{q}_{ab} - 2 D_{(a}N_{b)}\right)
\end{equation}
in terms of a spatial covariant derivative $D_a$. The lapse function
$N$ and shift vector $N^a$ do not have momenta and are not dynamical
since they do not appear as time derivatives in the action. They
rather play the role of multipliers to constraints which will be
written explicitly below for the vector mode.

However, in view of including quantum gravity effects we do not use
metric variables but connection variables which follow after a
canonical transformation and provide the basis for a loop quantization
of gravity \cite{LoopRep}.  In these Ashtekar variables
\cite{AshVar,AshVarReell} one uses a densitized triad $E^a_i$ instead
of the spatial metric, defined as
\begin{equation} \label{DTriad}
E^a_i := |\det (e_a^i)| e^a_i
\end{equation}
in terms of the co-triad $e_a^i$ and its inverse $e^a_i$ which in turn
are defined via the spatial metric as $q_{ab} =: e_a^i e_b^i$. The
canonically conjugate variable to the densitized triad is the Ashtekar
connection $A_a^i := \Gamma_a^i+\gamma K_a^i$, where $K_a^i$ is
extrinsic curvature and $\gamma>0$ is the so-called Barbero-Immirzi
parameter \cite{AshVarReell,Immirzi}. The spin connection $\Gamma_a^i$
by definition satisfies $D_ae^a_i=0$ which can be solved as
\begin{equation} \label{SpinConnection}
 \Gamma_a^i= -\epsilon^{ijk}e^b_j (\partial_{[a}e_{b]}^k+
 {\textstyle\frac{1}{2}} e_k^ce_a^l\partial_{[c}e_{b]}^l)\,
\end{equation}
%
%
As mentioned, we perturb basic variables around a spatially-flat
FRW background. Further, we denote background variables with a bar
as in
\begin{equation} \label{BGVariables}
\bar{E}^a_i = \bar{p} \delta^a_i \quad,\quad
\bar{\Gamma}_a^i = 0 \quad,\quad \bar{K}_a^i = \bar{k}\delta_a^i 
\quad,\quad \bar{N} = \sqrt{\bar{p}} \quad,\quad \bar{N}^a = 0 ~,
\end{equation}
where $\bar{p} = a^2$ and the spatial metric is $\bar{q}_{ab} =
a^2\delta_{ab}$. (In general, $\bar{p}$ as a triad rather than metric
component could be negative, which however we can safely ignore here.)
The choice of $\bar{N}=a$ leads to conformal time. One may emphasize
here that for a spatially flat FRW spacetime the background spin
connection $\bar{\Gamma}_a^i$ vanishes, thus $\bar{A}_a^i= \gamma
\bar{K}_a^i= \gamma\bar{k} \delta_a^i$. Moreover, $\bar{k}= {\rm
  d}a/{\rm d}\eta= \dot{a}/a$ in conformal time.

\subsection{Perturbed canonical variables}

The perturbed densitized triad and Ashtekar connection around a
spatially flat FRW background are given by
\begin{equation} \label{PCVariables}
E^a_i = \bar{p} \delta^a_i + \delta E^a_i \quad,\quad
A_a^i = \Gamma_a^i+\gamma K_a^i = \gamma\bar{k}\delta_a^i 
+ (\delta\Gamma_a^i+\gamma\delta K_a^i) ~,
\end{equation}
where $\bar{p}$ and $\gamma\bar{k}$ are the background densitized
triad and Ashtekar connection. At the linear level, the general
solution for the co-triad corresponding to $q_{ab}=
a^2(\delta_{ab}+F_a,_b+F_b,_a)$ as in equation (\ref{PMVector}) is
\begin{equation} \label{CTVector}
e_a^i = a\left[\delta_a^i+ (c_1 {F_a,}^i + c_2 {F^i},_{a})\right] ~, 
\end{equation}
where $c_1+c_2=1$. Specific values of $c_1$ and $c_2$ are part of the
gauge choice of the triad as a set of three vector fields in arbitrary
rotation. Using definition (\ref{DTriad}) for the densitized triad one
can easily compute the expression of the perturbed densitized triad for
vector mode
\begin{equation} \label{PDTVector}
\delta E^a_i = -\bar{p}(c_1 {F_i,}^a + c_2 {F^a},_{i}) ~, 
\end{equation}
where we have used the divergence free property, i.e.\
$\delta_a^i\delta E^a_i=0$, for vector mode perturbations; thus no
linear term results from the determinant used in defining the
densitized triad. By comparing the perturbed spacetime metric from
expression (\ref{PMVector}) with the relation (\ref{MetricRelation}),
one can read off the expression for the perturbed lapse function and
shift vector
\begin{equation} \label{PerturbedLapseShift}
\delta N = 0  \quad,\quad \delta N^a = S^a  ~. 
\end{equation}
Using the spin connection (\ref{SpinConnection}) and the
general expression of the perturbed densitized triad
in (\ref{PCVariables}), the linearized spin connection is
given by
\begin{equation} \label{PSpinConnection}
\delta \Gamma_a^i = \frac{1}{\bar{p}} \epsilon^{ije} \delta_{ac} 
\partial_e \delta E^c_j  ~. 
\end{equation}

\subsection{Canonical structure of linearized vector modes}

In a canonical formulation, the Einstein-Hilbert action can be written
equivalently (up to boundary terms) using the Ashtekar connection and
densitized triad as
\begin{equation} \label{EHAction}
S_{EH} = \int {\rm d}t \int_{\Sigma} \mathrm{d}^3x
\left[ \frac{1}{8\pi G\gamma} E^a_i {\mathcal L}_t A_a^i - 
\left[\Lambda^i G_i+N^a{\mathcal C}_a+N{\mathcal C}\right]\right]~,
\end{equation}
where $\Lambda^i$, $N^a$ and $N$ are Lagrange multipliers of the
Gauss, diffeomorphism and Hamiltonian constraints, explicit
expressions of which are written below. Before decomposing the
symplectic structure according to (\ref{PCVariables}), we introduce a
cell to render the homogeneous mode well-defined. Integrating the
first term of (\ref{EHAction}) only over a finite box of coordinate
volume $V_0$ with perturbed basic variables of the form
(\ref{PCVariables}), we obtain the symplectic structure given by the
Poisson brackets of the background and perturbed variables as
\begin{equation} \label{PoissonAlgebra}
\{\bar{k}, \bar{p} \} = \frac{8\pi G}{3 V_0} \quad,\quad
\{\delta K_a^i(x), \delta E^b_j(y)\} = {8\pi G} \delta^3(x,y) \delta_a^b
\delta^i_j ~.
\end{equation}
In deriving these relation, we have used the properties that for
vector perturbations $\delta_a^i \delta E^a_i = \delta^a_i \delta
K_a^i = 0$. This provides separate canonical structures for the
background and perturbations, but these variables will be coupled
dynamically. In particular, the homogeneous background is dynamical
and would receive back-reaction effects at higher than linear orders.

\subsubsection{Gauss constraint}

In triad variables, a Gauss constraint appears which generates
internal gauge rotations of phase space functions because triads whose
legs are rotated at a fixed point correspond to the same spatial
metric. This constraint is given by
\begin{equation} \label{GaussConstraint}
G(\Lambda) := \int_{\Sigma} \mathrm{d}^3x\Lambda^i G_i
= \frac{1}{8\pi G\gamma}\int_{\Sigma}\mathrm{d}^3x\Lambda^i
\left(\partial_aE^a_i+{\epsilon_{ij}}^k A_a^j E^a_k\right) ~.
\end{equation}
Using the perturbed form of basic variables (\ref{PCVariables}), it
can be reduced to
\begin{equation} \label{PertGaussConstraint}
G(\Lambda) = \frac{1}{8\pi G}\int_{\Sigma}\mathrm{d}^3x\Lambda^i
\left({\epsilon_{ij}}^a \bar{p}\delta K_a^j 
+ {\epsilon_{ia}}^k \bar{k} \delta E^a_k\right) ~.
\end{equation}
Since we are working with a background $E^a_i=\bar{p}\delta_i^a$ whose
gauge freedom is fixed, the multiplier $\Lambda^i$ is already of first
order. To derive the expression (\ref{PertGaussConstraint}), we have
used the definition of the spin connection $\Gamma_a^i$ in terms of
the densitized triad $E^a_i$, which reduces to (\ref{PSpinConnection})
for the linearized equations. Internal gauge rotations of phase space
functions are parametrized by the Lagrange multiplier $\Lambda^i$
through $\delta_{\Lambda} f = \{f, G[\Lambda]\}$. In particular, the
internal gauge rotations of perturbed basic variables are
\begin{equation} \label{GaugeRoration}
\delta_{\Lambda}(\delta K_a^i) := \{\delta K_a^i, G(\Lambda)\} 
= \bar{k} \Lambda^l {\epsilon_{la}}^i \quad,\quad
\delta_{\Lambda}(\delta E^a_i) := \{\delta E^a_i, G(\Lambda)\} 
= -\bar{p} \Lambda^l {\epsilon_{li}}^a ~.
\end{equation}
Clearly, the perturbed variables themselves are not invariant under
internal gauge rotations in spite of the fixed background. However,
one may notice already that the symmetrized perturbed variables are in
fact invariant under internal gauge rotations. Thus, the physical
quantities must depend only on the symmetrized form of the perturbed
basic variables. Then, the constants $c_1$ and $c_2$ in
(\ref{PDTVector}) are irrelevant since
\[
 E_{(i}{}^{a)} = -\frac{1}{2}\bar{p}(F_i,{}^a+F^a{},_i)\,.
\]

\subsubsection{Diffeomorphism constraint}

The diffeomorphism constraint generates gauge transformations
corresponding to spatial coordinate transformations of phase space
functions. Its general contribution from gravitational variables is
given by
\begin{equation} \label{DiffeoConstraint}
D_G[N^a] := \int_{\Sigma} \mathrm{d}^3xN^a{\mathcal C}_a
= \frac{1}{8\pi G\gamma}\int_{\Sigma}\mathrm{d}^3xN^a
\left[ F_{ab}^i E^b_i - A_a^i G_i \right]
\end{equation}
(where the subscript ``G'' stands for ``gravity'' to separate the term
from the matter contribution.  In general, a matter field would also
contribute a term $D_m$ to the diffeomorphism constraint, which we
leave unspecified here and relate later to the stress-energy tensor.)
The second term vanishes by virtue of the Gauss constraint, but is
necessary to generate diffeomorphisms in the form of Lie derivatives
of phase space functions along the shift vector.  Using the expression
of the perturbed basic variables (\ref{PCVariables}), one can reduce
the diffeomorphism constraint to
\begin{equation} \label{ClassPertDiffConst}
D_G[N^a] = \frac{1}{8\pi G}\int_{\Sigma}\mathrm{d}^3x\delta N^c
\left[ -\bar{p}(\partial_k\delta K^k_c) - \bar{k} \delta_c^k(
\partial_d \delta E^d_k)\right] ~.
\end{equation}
Here, we have kept up to
quadratic terms in the perturbations, noting $\bar{N}^a=0$.

\subsubsection{Hamiltonian constraint}

In a canonical formulation, the Hamiltonian constraint generates `time
evolution' of the spatial manifold for phase space functions
satisfying the equations of motion. Its gravitational contribution in
Ashtekar variables is
\begin{equation}\label{HamConstraint}
 H_G[N] = \frac{1}{16\pi G} \int_{\Sigma} \mathrm{d}^3x N 
\frac{ E^c_jE^d_k}{\sqrt{\left|\det E\right|}}
\left({\epsilon_i}^{jk}F_{cd}^i -2(1+\gamma^{2}) 
K_{[c}^j K_{d]}^k\right) ~.
\end{equation}
Using the general perturbed forms of basic variables and the
expression of curvature $F_{ab}^i = \partial_a A_b^i - \partial_b
A_a^i + {\epsilon^i}_{jk} A_a^j A_b^k$, one can simplify
(\ref{HamConstraint}). Up to quadratic terms it is given by
\begin{eqnarray} \label{ClassPertHamConst}
H_G[N] = \frac{1}{16\pi G}\int_{\Sigma}\mathrm{d}^3x \bar{N}
\left[-6\bar{k}^2\sqrt{\bar{p}} - 
 \frac{\bar{k}^2}{2\bar{p}^\frac{3}{2}}\right. 
(\delta E^c_j\delta E^d_k\delta_c^k\delta_d^j) \nonumber\\ +
\left.\sqrt{\bar{p}} (\delta K_c^j\delta K_d^k\delta^c_k\delta^d_j)
 - \frac{2\bar{k}}{\sqrt{\bar{p}}} (\delta E^c_j\delta K_c^j)
\right] 
\end{eqnarray}
with $\delta N=0$ for vector modes.  We may mention here that $\gamma$
dependent terms drop out of the Hamiltonian constraint once the spin
connection, the generic form (\ref{PDTVector}) of the densitized triad
and the extrinsic curvature are used. Again, there is a matter
contribution $H_m$ left unspecified here.

\subsection{Gauge transformations and gauge invariant variables}

General gauge transformations are determined by a choice of $N$ and
$N^a$, which gives rise to all space-time coordinate transformations.
For the vector mode, only the choice $N^a=\xi^a$ with $\xi^a,_a=0$ is
relevant since the remaining functions $N$ or $N_a=\xi,_a$ would
affect only the scalar mode. Thus, transformations of interest here
are generated only by diffeomorphism constraint in a form
parametrized by the shift vector as $\delta_{\xi} f = \{f,
D_G[N^a=\xi^a]\}$. We need to consider this only to linear order,
using the linearized constraints. The resulting transformation for
basic variables will then be used to find gauge invariant
combinations. Alternatively, a canonical formulation allows one to
compute gauge invariant observables first, and then linearize
\cite{PertObsI,PertObsII}.

Under such a gauge transformation, the perturbed
densitized triad and extrinsic curvature transform as
\begin{equation} \label{GTofdEndK} 
\delta_{\xi}(\delta E^a_i) = -\bar{p} \partial_i \xi^a \quad,\quad
\delta_{\xi}(\delta K_a^i) = \bar{k} \partial_a \xi^i ~.
\end{equation}
Relating $\delta E^a_i$ to $\delta F^a$ and $\delta K_a^i$ to $\delta
S^a$ based on (\ref{Kab}) leads to gauge transformations for the vector
mode functions $F^a$ and $S^a$:
\begin{equation} \label{GTFnS} 
\delta_{\xi}F^a = \xi^a \quad,\quad
\delta_{\xi}S^a = \dot{\xi}^a \quad,\quad
\delta_{\xi}\sigma^a = \delta_{\xi}(S^a-\dot{F}^a) = 0
\end{equation}
introducing $\sigma^a:=S^a-\dot{F}^a$ as a gauge invariant variable
for the gravitational vector mode.

\subsection{Linearized equations of motion}

General equations of motion are written canonically as
\begin{equation} \label{eom}
 \dot{f} = \{f,{\cal H}\}
\end{equation}
for any phase space function $f$ using the total Hamiltonian ${\cal
  H}$. For gravity, ${\cal H}= H_G[N]+D_G[N^c]+H_{m}[N]+D_{m}[N^c]$
with the gravitational contributions $H_G$ and $D_G$ and matter
contributions $H_m$ and $D_m$ to the Hamiltonian and diffeomorphism
constraints. Equations of motion refer to coordinate time, with
derivatives being indicated by the dot. The form of the lapse function
$N$ specifies which time coordinate is used; here,
$\bar{N}=a=\sqrt{\bar{p}}$ implies conformal time $\eta$. The general
form (\ref{eom}) also applies to equations of motion for momenta of
the multipliers, such as $\dot{P}_N=0=\{P_N,{\cal H}\}= -\delta {\cal
  H}/\delta N$. This must be zero because the momentum $P_N$ is zero
for an action (\ref{EHAction}) not depending on the time derivative of
$N$. In this way, equations of motion for the momenta of lapse and
shift give rise to constraints.

Hamilton's equation of motion for the perturbed densitized triad
\begin{equation} 
\delta\dot{E}^a_i = 
\{\delta E^a_i, {\cal H}\}
\end{equation}
leads to the expression (\ref{Kab}) of extrinsic curvature, but
linearized.  We will only need it in symmetrized form which is
\begin{equation} \label{PECVector}
\delta {K_{(a}}^{i)} =  
\frac{1}{2}\left[ \bar{k}({F_a,}^i + {F^i},_{a})
+ {({\dot{F}_a,}^i+{\dot{F}}^i},_{a})-({S_a,}^i + {S^i},_{a}) \right] ~.
\end{equation}
or
\begin{equation} \label{PertExtCurvature}
\delta {K_{(a}}^{i)} = -\frac{1}{2}({\sigma_a,}^i+{\sigma^i},_{a}) + 
\frac{1}{2}\bar{k}({F_a,}^i + {F^i},_{a})
\end{equation}
using the gauge-invariant variable $\sigma^a$.
Variation with respect to the shift vector $\delta N^a$,
\begin{equation} \label{ClassSigmaLaplacianEqn0}
\frac{\delta}{\delta (\delta N^a)}{\cal H} = 0\,, 
\end{equation}
when expressed in terms of symmetrized vector perturbations implies
\begin{equation} \label{ClassSigmaLaplacianEqn}
-\frac{\bar{p}}{2} ~\nabla^2 \sigma_a ~=~ 8\pi G
\left[\frac{\delta D_{m}}{\delta (\delta N^a)}\right] ~.
\end{equation}
Using the relation between the perturbed stress-energy tensor
and a variation of the matter diffeomorphism constraint $D_m$ with respect
to the shift vector as derived in \cite{HamPerturb}, 
\begin{equation} \label{DiffeoVarSETRelation}
- \frac{1}{\bar{N}\bar{p}^{3/2}} 
\left[\frac{\delta D_{m}}{\delta (\delta N^a)}\right]
= \delta {{T^{(v)}}^{0}}_{a} 
=:- (\rho+P) V_a ~.
\end{equation}
one can express
equation (\ref{ClassSigmaLaplacianEqn}) in standard form
\cite{VectorColl} 
\begin{equation} \label{ClassSigmaLaplacianEqnBB}
- \frac{1}{2a^2} ~\nabla^2 \sigma_a ~=~ 8\pi G
(\rho+P) V_a
\end{equation}
for a vector mode equation.  Here $\rho$ and $P$ are energy density
and pressure of the background matter field.

The second vector mode equation comes from Hamilton's
equation of motion for perturbed extrinsic curvature, which using
(\ref{PertExtCurvature}) becomes
\begin{equation} \label{ClassSigmadotEqn}
-\frac{1}{2}\frac{{\rm d}}{{\rm d}\eta}({\sigma_a,}^i+{\sigma^i},_{a}) 
-\bar{k} ({\sigma_a,}^i+{\sigma^i},_{a}) =
8\pi G \bar{p} \delta {{T^{(v)}}^{(i}}_{a)} ~. 
\end{equation}
The perturbed spatial stress tensor for the vector mode in terms of
the matter Hamiltonian \cite{HamPerturb} is
\begin{equation} \label{ClassStressTensor}
\delta {{T^{(v)}}^{(i}}_{a)} =
\frac{1}{\bar{p}} 
\left[\frac{1}{3V_0}\frac{\partial H_{m}}{\partial\bar{p}} 
\left(\frac{\delta E^c_j\delta^j_{(a}\delta^{i)}_c}{\bar{p}}\right) +
\frac{\delta H_{m}}{\delta(\delta {E^{(a}}_{i)})}\right] \,.
\end{equation}
By using the expression of the perturbed stress-energy tensor in terms
of anisotropic stress $\pi_a$ as $\delta {{T^{(v)}}^{(i}}_{a)} =: P
({\pi_a,}^i+{\pi^i},_{a})$ we can express equation
(\ref{ClassSigmadotEqn}) in standard form \cite{VectorColl}
\begin{equation} \label{ClassSigmadotEqnBB}
-\frac{1}{2a^4}\frac{{\rm d}}{{\rm d}\eta}
\left[ a^2 ({\sigma_a,}^i+{\sigma^i},_{a})\right] 
= 8\pi G P({\pi_a,}^i+{\pi^i},_{a}) 
\end{equation}
for the second vector mode equation.

For vector modes, variation with respect to the lapse function does
not give new field equations but would rather contribute back-reaction
terms to the background evolution.
The two equations (\ref{ClassSigmaLaplacianEqn}) and
(\ref{ClassSigmadotEqn}) thus provide the complete dynamics.

\section{Quantum dynamics: Inverse triad corrections}

We have completed the derivation of vector mode equations in
Hamiltonian cosmology based on Ashtekar's formulation of general
relativity. 
This naturally agrees with results of \cite{VectorColl}.
As mentioned, we are interested in applying this
formulation to study possible canonical quantum gravity effects.  As
an explicit example, we now consider quantum corrections coming from
terms containing inverse densitized triad components.  In
Sec.~\ref{S:Hol} we provide formulas for a second major quantum
correction that one expects from the use of holonomies, rather than
direct connection components, as basic operators of the quantum
theory.  This has the effect of adding terms of higher order in
extrinsic curvature components, and thus higher powers of the first
time derivative of the metric, to the Hamiltonian.  In addition to
those two corrections, there are higher derivative corrections implied
by genuine quantum effects \cite{EffAc,Karpacz}.  All this combines to
effective constraints or effective equations of motion for the system.
As for the diffeomorphism constraint, we assume that it receives no
quantum corrections because it is quantized directly through its phase
space transformations \cite{ALMMT}.

\subsection{Quantum corrected Hamiltonian constraint}

While homogeneous quantum cosmology using loop quantum gravity
techniques is rather well understood \cite{LivRev}, a systematic
derivation of quantum corrections to classical dynamics which includes
inhomogeneity is not yet available. But typical effects are known and
provide valuable indications for implications of quantum effects. 
In loop quantum gravity, the
appearance of inverse powers of the densitized triad as in $\frac{
  E^c_jE^d_k}{\sqrt{\left|\det E\right|}}$ initially leads to
difficulties since flux operators quantizing the densitized triad have
discrete spectra containing zero as an eigenvalue \cite{AreaVol,Area}.
These difficulties can be overcome in a way exploiting background
independence, and giving rise to well-defined operators
\cite{QSDI,QSDV}. However, for small values of the densitized triad,
where the classical expression would diverge, there are deviations
from the classical behavior which imply quantum corrections. For the
homogeneous case, explicit calculations show that the classical term
is multiplied by a factor $\bar{\alpha}$ \cite{InvScale} making
the whole expression finite. For large fluxes, the leading terms in
explicit expressions of $\bar{\alpha}$ are of the form
\begin{equation} \label{Alphafunction}
\bar{\alpha}(\bar{p}) = 1 + c {\left(\frac{\ell_{\rm P}^2}{\bar{p}}\right)}^n ~,
\end{equation}
where $n$ and $c$ are positive definite numbers which correspond to a
given inverse triad operator. However, they are not completely fixed
since triad operators themselves are subject to quantization
ambiguities \cite{Ambig,ICGC}. One important motivation to study
inhomogeneous models is that, compared to homogeneous models, their
dynamics gives rise to much tighter consistency conditions which could
constrain such parameters. This is indeed borne out, as we will see
later. While general derivations of $\alpha$ for inhomogeneous
configurations without assumptions on the metric such as symmetries or
specific modes is complicated (see e.g.\ \cite{BoundFull}) and does
not yet provide many insights, $\alpha$ can be computed in certain
perturbative regimes. It has been studied recently for scalar mode
perturbations \cite{HamPerturb}, showing a similar behavior as in the
isotropic case. Here we consider such quantum corrections for vector
mode perturbations, although this is a case where no explicit
expression is available yet. As we will see, consistency itself
restricts the form of $\alpha$ beyond what can currently be computed
directly from operators. Thus, our perturbative treatment at the
effective level provides conjectures to probe the overall consistency
of the theory by comparing with results for the underlying operators.

This procedure leads us to an ansatz for the quantum corrected
Hamiltonian constraint
\begin{eqnarray} \label{QMPertHamConst}
H_G^Q[N] = \frac{1}{16\pi G}\int_{\Sigma}\mathrm{d}^3x \bar{N} 
\alpha(\bar{p},\delta E^a_i)
\left[-6\bar{k}^2\sqrt{\bar{p}} - \frac{\bar{k}^2}{2\bar{p}^\frac{3}{2}}\right. 
(\delta E^c_j\delta E^d_k\delta_c^k\delta_d^j) \nonumber\\
+ \left.\sqrt{\bar{p}} (\delta K_c^j\delta K_d^k\delta^c_k\delta^d_j) 
- \frac{2\bar{k}}{\sqrt{\bar{p}}} (\delta E^c_j\delta K_c^j) 
\right] ~,
\end{eqnarray}
where $\alpha(\bar{p},\delta E^a_i)$ is the correction function, now
also depends on triad perturbations. It is important to emphasize
here that the correction $\alpha$ coming from the quantized inverse
densitized triad in general could be tensorial in nature. However,
later we will see that the leading effect on perturbation dynamics
comes from the background corrections, i.e.\ from
$\bar{\alpha}=\alpha(\bar{p},0)$, as well as derivatives of $\alpha$
evaluated at the background configuration. (Note that the only
background variable determining the geometry is $\bar{p}$, in which
the corrections are expressed. This is sometimes seen as problematic
since the scale factor $a=\sqrt{\bar{p}}$ can be rescaled arbitrarily
in a flat isotropic model. However, the dependence of a function
$\alpha$ in an inhomogeneous Hamiltonian constraint is through
elementary fluxes whose values are determined by an underlying
inhomogeneous state. The scale of corrections, too, is determined by
the underlying state, resolving any apparent contradiction between the
appearance of such a scale and the rescaling freedom of a flat,
precisely isotropic background.)

\subsection{Constraint algebra}

In a canonical formulation of general relativity, classical
constraints form a first class Poisson algebra, i.e.\ $\{C_I,C_J\}=
f^{K}_{IJ}(A,E) C_K$, whose coefficients can in general be structure
functions. It ensures that the transformations generated by the
constraints are gauge and preserve the constraint surface. In other
words, the evolution of phase space functions preserves the physical
solution surface. To study quantum gravity effects, we have introduced
a quantum correction function $\alpha(\bar{p},\delta E^a_i)$ which
depends on phase space variables.  Naturally, having a new expression
for the Hamiltonian constraint, there could be an anomaly term of
quantum origin in the constraint algebra. Here, while
$\{H_G^Q[N],H_G^Q[N']\}$ is trivial in the absence of lapse
perturbations for the vector mode, a non-trivial anomaly in the
algebra could occur in the Poisson bracket between $H_G^Q[N]$ and
$D_G[N^a]$. This bracket turns out to be
\begin{equation} \label{HDAnomaly}
\{H_G^Q[N], D_G[N^a]\} = 
\frac{1}{8\pi G}\int_{\Sigma}\mathrm{d}^3x
\bar{p} (\partial_j \delta N^c){{\mathcal A}_c}^j ~,
\end{equation}
where
\begin{equation} \label{AnomalyExpr}
{{\mathcal A}_c}^j = 
3 \bar{N}\bar{k}^2 \sqrt{\bar{p}}
\left[\frac{\partial \alpha}{\partial (\delta E^c_j)} 
+ \frac{1}{3 \bar{p}} \frac{\partial\alpha}{\partial\bar{p}} 
\left(\delta E^d_k\delta^k_c\delta^j_d\right)
\right] ~.
\end{equation}
We see that the anomaly term contains derivative of $\alpha$ with
respect to both $\bar{p}$ and $\delta E^a_i$. However, as mentioned
before the functional form of the correction function
$\alpha(\bar{p},\delta E^a_i)$ in terms of $\delta E^a_i$ is not known
while the $\bar{p}$-dependence can be taken to be of scalar mode form,
i.e.\ (\ref{Alphafunction}) with parameters $c$ and $n$ fixed once an
inverse triad operator is chosen. To have a consistent set of
evolution equations for the vector mode we require the anomaly term to
vanish i.e.\ ${{\mathcal A}_c}^j=0$. This in turn puts restrictions on
the linearized functional form of $\alpha$ as a function of $\delta
E^a_i$:
\[
 \frac{\partial \alpha}{\partial (\delta E^c_j)} 
=- \frac{1}{3 \bar{p}} \frac{\partial\alpha}{\partial\bar{p}} 
\left(\delta E^d_k\delta^k_c\delta^j_d\right)\,.
\]
Since $\alpha$ is in principle computable in the full theory, this
provides important consistency checks for loop quantum gravity.  At
present, only the dependence $\alpha(\bar{p})$ as well as derivatives
of $\alpha$ along diagonal components of the spatial metric are known
\cite{QuantCorrPert}.  Anomaly cancellation will then lead us to
conjecture a form of derivatives $\delta\alpha/\delta (\delta E_i^a)$
along off-diagonal components of the metric which one can later
compare with direct calculations once they become available.
 
\subsection{Effective gauge invariant perturbation and its linearized
  equation of motion}

Extrinsic curvature is derived using one of Hamilton's equations of
motion. Thus, one expects the expression of extrinsic curvature as it
follows from an equation of motion to change due to the quantum
corrections. This incorporates an effect of quantum geometry which
changes the usual differential geometric relation of extrinsic
curvature corresponding to changes of the spatial metric between
different slices. One can easily compute the corrected expression for
the perturbed extrinsic curvature
\begin{equation} \label{QMPertExtCurvature}
\delta {K_{(a}}^{i)} = -\frac{1}{2\alpha}
({\sigma_a,}^i+{\sigma^i},_{a}) +
\frac{1}{2}\bar{k}({F_a,}^i + {F^i},_{a})
\end{equation}
where $\alpha$ appears only in the first term expressed through the
classical gauge invariant quantity $\sigma^a=S^a-\dot{F}^a$.  

The gauge transformation of the vector perturbation $F^a$ remains
unchanged because the diffeomorphism constraint retains its classical
form. Even though there is a quantum correction to extrinsic
curvature one can easily see that $\sigma^a = S^a-{\dot{F}}^a$ is
still the gauge invariant variable.  Moreover with the diffeomorphism
constraint being unaffected, Eq.~(\ref{ClassSigmaLaplacianEqn})
remains unchanged for the quantum background dynamics.  Now using the
second Hamilton's equation, one obtains an equation of motion for
extrinsic curvature as follows
\begin{equation} \label{QMSigmadotEqn}
  \frac{1}{\bar{\alpha}}\left[-\frac{1}{2}\frac{{\rm d}}{{\rm
        d}\eta}({\sigma_a,}^i+{\sigma^i},_{a})
    -\bar{k}(\bar{\alpha}-{\bar{\alpha}}^{'}\bar{p})
    ({\sigma_a,}^i+{\sigma^i},_{a}) \right] + ~{{\mathcal
      A}_{(a}}^{i)} = 8\pi G \bar{p} \delta {{T^{(v)}}^{(i}}_{a)} ~~.
\end{equation}
To derive Eq.~(\ref{QMSigmadotEqn}), we have used the corrected
expression (\ref{QMPertExtCurvature}) of extrinsic curvature as well
as an analogous expression for the background extrinsic curvature,
$\bar{\alpha}\bar{k}=\dot{a}/a$. One may note here that equation
(\ref{QMSigmadotEqn}) explicitly contain the anomaly term.  Thus, the
requirement of an anomaly free constraint algebra leads to the
corrected equation (\ref{QMSigmadotEqn}) explicitly in terms of the
gauge invariant variable. The presence of an anomaly, on the other
hand, would make it impossible to express the equations of motion
solely in terms of gauge invariant variables. Since consistency of the
constraint algebra requires us to set ${\cal A}_a^i=0$, closed
equations for the gauge invariant perturbations follow. Nevertheless,
non-trivial quantum corrections remain through $\bar{\alpha}$.

Note that the only correction function $\alpha$ in the Hamiltonian
appears in a form multiplying the lapse function $N$. The correction
could thus be reduced to a simple change of the lapse function and
thus the time gauge. Still, the corrections are non-trivial as
illustrated by the equations of motion shown here. The choice and
interpretation of time is based on the line element since this
determines the measurement process of co-moving geodesic observers.
This behavior is not changed by the appearance of a correction
function $\alpha$ in the Hamiltonian constraint even if it always
appears in combination with $N$. Even with a corrected Hamiltonian
constraint we are still referring to the same form of conformal time,
but fields evolve differently as given by the corrected constraint.
Then, also observable implications of the quantum corrections are
possible.

\section{Quantum dynamics: Holonomy corrections}
\label{S:Hol}

In addition to corrections in coefficients of the constraint due to
inverse powers of densitized triad components, there are corrections
which resemble higher curvature terms in an effective action. While
these corrections would be dominant in purely isotropic models by
virtue of the large matter energy density in a macroscopic universe,
they are sub-dominant in inhomogeneous situations \cite{InhomLattice}.
Moreover, as we will see they do not provide much of a structural
change to the equations. The Hamiltonian constraint operator is
formulated in terms of holonomies rather than connection or extrinsic
curvature components. Since these objects are non-linear as well as
(spatially) non-local in connection components, they provide higher
order and higher spatial derivative terms. Higher time derivatives, as
they would also be provided by higher curvature terms, do not arise in
this way but rather through the coupling of fluctuations and higher
moments of a quantum state to the expectation values
\cite{EffAc,Karpacz}. The full effective constraint including all
these terms has not been derived yet. In this section we therefore
focus on an analysis of higher order terms only.

For an isotropic model sourced by a massless, free scalar field such
higher order terms turn out to be the only corrections, and there are
no higher time derivatives \cite{BouncePert}.  The exact effective
Hamiltonian can then be obtained by simply replacing the background
Ashtekar connection $\gamma\bar{k}$ by
$\bar{\mu}^{-1}\sin\bar{\mu}\gamma\bar{k}$, as it was also seen in
numerical studies \cite{APS}. The parameter $\bar{\mu}$ depends on the
quantization scheme and may be a function of $\bar{p}$. Just as with
the parameter $c$ in (\ref{Alphafunction}), we will see that the
freedom is constrained by anomaly cancellation. To study the effects
of the background dynamics on inhomogeneous perturbations, we
similarly substitute the appearance of $\bar{k}$ in the classical
Hamiltonian by a general form $(m\bar{\mu})^{-1}\sin
m\bar{\mu}\gamma\bar{k}$ where $m$ is an integer. (This parameter is
kept free because different factors of sines and cosines combine from
the full constraint to result in this term. It can be constrained by
looking at detailed properties of the underlying operator, but also by
consistency requirements as we will see shortly.) With this
prescription, one can write down expression for the corrected
Hamiltonian constraint
\begin{eqnarray} \label{QMPertHamConstHolo}
H_G^Q[N] = \frac{1}{16\pi G}\int_{\Sigma}\mathrm{d}^3x \bar{N} 
\left[-6\sqrt{\bar{p}}
\left(\frac{\sin\bar{\mu}\gamma\bar{k}}{\bar{\mu}\gamma}\right)^2
 - \frac{1}{2\bar{p}^\frac{3}{2}} 
\left(\frac{\sin\bar{\mu}\gamma\bar{k}}{\bar{\mu}\gamma}\right)^2
(\delta E^c_j\delta E^d_k\delta_c^k\delta_d^j) \right. \nonumber\\
+ \left.\sqrt{\bar{p}} (\delta K_c^j\delta K_d^k\delta^c_k\delta^d_j) 
- \frac{2}{\sqrt{\bar{p}}} 
\left(\frac{\sin 2\bar{\mu}\gamma\bar{k}}{2\bar{\mu}\gamma}\right)
(\delta E^c_j\delta K_c^j) 
\right] ~.
\end{eqnarray}
Here we have required that the effective Hamiltonian
(\ref{QMPertHamConstHolo}) has a homogeneous limit in agreement with
what has been used in isotropic models. This fixes the parameter $m$
to equal one in the first two terms. The parameter for the last term
as chosen here is the one which leads to an anomaly-free constraint
algebra.

Although we write explicit sines in this expression, and thus
arbitrarily high powers of curvature components, it is to be
understood only as a short form to write the leading order
corrections. Higher orders are supplemented by further, yet to be
computed higher curvature quantum corrections. The expressions are
thus reliable only when the argument of the sines is small, which
excludes the bounce phase itself. (Such sine corrections can be used
throughout the bounce phase only for exactly isotropic models sourced
by a free, massless scalar \cite{BouncePert,BounceCohStates}, but not
in the presence of a matter perturbation \cite{BouncePot} or
anisotropies and inhomogeneities.)

\subsection{Constraint algebra}

Again, a non-trivial anomaly in the algebra can occur between the
Poisson bracket between $H_G^Q[N]$ and $D_G[N^a]$
\begin{equation} \label{HDAnomalyHolo}
\{H_G^Q[N], D_G[N^a]\} = \frac{\bar{N}}{\sqrt{\bar{p}}} 
\left(\bar{k} - \frac{\sin 2\bar{\mu}\gamma\bar{k}}
{2\bar{\mu}\gamma}\right) D_G[N^a]
+ \frac{1}{8\pi G}\int_{\Sigma}\mathrm{d}^3x
\bar{p} (\partial_c \delta N^j){\mathcal A}_j^c ~,
\end{equation}
where
\begin{equation} \label{AnomalyExprHolo}
{\mathcal A}_j^c = \frac{\bar{N}}{\sqrt{\bar{p}}} 
\left[\bar{p}\frac{\partial}{\partial \bar{p}} 
\left(\frac{\sin\bar{\mu}\gamma\bar{k}}{\bar{\mu}\gamma}\right)^2
+ \left(\frac{\sin\bar{\mu}\gamma\bar{k}}{\bar{\mu}\gamma}\right)^2
 - \bar{k}^2 \right]
\left(\frac{\delta E^c_j}{\bar{p}}\right)
 ~.
\end{equation}
One may easily check here that $\bar{\mu}\sim 1/{\sqrt{\bar{p}}}$
leads to an anomaly free algebra up to order $\bar{k}^4$. 
This is in accordance with the result of arguments put forward recently
in purely isotropic models \cite{APSII}. From an inhomogeneous
perspective, the behavior $\bar{\mu}\sim 1/{\sqrt{\bar{p}}}$ reflects
the fact that the fundamental Hamiltonian creates new vertices when
acting on a graph state such that the number of vertices increases
linearly with volume \cite{InhomLattice,SchwarzN}.  This suggests a tight
relation between anomaly freedom at the effective level and properties
such as the creation of new vertices by a fundamental Hamiltonian
constraint.
 
\subsection{Effective linearized equation}


Hamilton's equations governing the background dynamics are given
by
\begin{equation} \label{BGDTEoMHolo}
{\dot{\bar{p}}} = 2\bar{p} \left(\frac{\sin
2\bar{\mu}\gamma\bar{k}} {2\bar{\mu}\gamma}\right)
\end{equation}
and
\begin{equation} \label{BGECEoMHolo}
{\dot{\bar{k}}} = -\frac{\bar{N}}{\sqrt{\bar{p}}} 
\left[\frac{1}{2}\left(\frac{\sin\bar{\mu}\gamma\bar{k}}
{\bar{\mu}\gamma}\right)^2 +\bar{p}\frac{\partial}{\partial \bar{p}} 
\left(\frac{\sin\bar{\mu}\gamma\bar{k}}{\bar{\mu}\gamma}\right)^2
\right] +\frac{8\pi G}{3 V_0}
\left(\frac{\partial\bar{H}_m}{\partial\bar{p}}\right)~.
\end{equation}
Since the diffeomorphism constraint is assumed to remain unaffected,
the gauge transformation of vector functions $F^a$ remains unchanged.
However, a correction to the extrinsic curvature expression now leads
to a new expression for the gauge invariant variable
\begin{equation} \label{GIVarHolo}
\sigma^a = S^a-{\dot{F}}^a + \bar{k}\left(1 
-\frac{\sin 2\bar{\mu}\gamma\bar{k}} 
{2\bar{\mu}\gamma\bar{k}}\right) F^a ~.
\end{equation}
The expression of perturbed extrinsic curvature in
terms of the gauge invariable variable remains unchanged, though,
\begin{equation} \label{QMPertExtCurvatureHolo}
\delta {K_{(a}}^{i)} = -\frac{1}{2}
({\sigma_a,}^i+{\sigma^i},_{a}) +
\frac{1}{2}\bar{k}({F_a,}^i + {F^i},_{a}) ~~. 
\end{equation}

Now using again Hamilton's equation, one obtains an equation of
motion for extrinsic curvature as follows
\begin{equation} \label{QMSigmadotEqnHolo}
-\frac{1}{2}\frac{{\rm d}}{{\rm d}\eta}({\sigma_a,}^i+{\sigma^i},_{a}) 
- \frac{1}{2}\bar{k}\left(1 +\frac{\sin 2\bar{\mu}\gamma\bar{k}} 
{2\bar{\mu}\gamma\bar{k}}\right)({\sigma_a,}^i+{\sigma^i},_{a})
 + ~{{\mathcal A}_{(a}}^{i)} =
8\pi G \bar{p} \delta {{T^{(v)}}^{(i}}_{a)} ~~. 
\end{equation}
As before, an anomaly-free constraint algebra, requiring ${\cal A}_a^i=0$,
leads to a quantum corrected equation entirely in terms of the gauge
invariant variable.

\section{Rate of change of vector perturbations}

We now have equations that govern the dynamics of vector mode
perturbations including quantum corrections. For simplicity, we
consider the situation where anisotropic stress is absent, which is
the context of \cite{VectorColl}.

\subsection{Classical dynamics}

Let us recall the key feature of classical vector mode perturbations.
In the absence of anisotropic stress, the right hand side of equation
(\ref{ClassSigmadotEqn}) vanishes. The classical background extrinsic
curvature is related to the time derivative of the scale factor $a$ as
$\bar{k} = \dot{a}/a$. It is also convenient to decompose the
amplitude of perturbations in terms of their Fourier modes $\sigma^i_k$.
With these simplifications, equation (\ref{ClassSigmadotEqn}) leads to
the rate of change
\begin{equation} \label{ClassGrowthRate}
\frac{{\mathrm d} \log \sigma^i_k}{ {\mathrm d} \log a } = - 2 
\end{equation}
for Fourier modes. Thus, any vector mode grows as $\sigma^i_k \sim
a^{-2}$ in a contracting phase, and correspondingly decays in an
expanding phase. This is independent of the background matter content
except for the assumed absence of anisotropic stress.  Thus, a quantum
correction of the background matter sector alone would not help in
taming the growth of vector perturbations.  Intuitively, one expects
that it is required to have a modified gravity sector in order to have
any modification of the growth rate. In the next sub-section we
consider the dynamics in the presence of quantum corrections to the
gravitational Hamiltonian.

\subsection{Quantum corrections: Inverse triad}

As in the classical case, we consider the situation where anisotropic
stress is absent.  Then using the equation of motion
$\bar{\alpha}\bar{k} = \dot{a}/a$ for the background, one can write
Eq.~(\ref{QMSigmadotEqn}) for Fourier modes
$\sigma^i_k$ of vector perturbations in the form
\begin{equation} \label{QMGrowthRate}
\frac{{\mathrm d} \log \sigma^i_k}{ {\mathrm d} \log a } = 
- 2\left(\frac{\bar{\alpha} -{\bar{\alpha}}^{'}\bar{p}}{\bar{\alpha}}\right) ~.
\end{equation}
For $\bar{\alpha}=1$ quantum corrections are switched off and we
obtain the classical result.
Now using the generic form of $\bar{\alpha}$ as in
(\ref{Alphafunction}) with an approach to one from above (i.e.\
$c>0$), it is easy to see that
\begin{equation} \label{QMGrowthRate2}
\left(\frac{\bar{\alpha} -{\bar{\alpha}}^{'}\bar{p}}{\bar{\alpha}}\right)
= 1 + n c {\left(\frac{\ell_{\rm P}^2}{\bar{p}}\right)}^n > 1 ~.
\end{equation}
Thus, the decay rate of vector mode perturbations is slightly higher
compared to that of a classical scenario. In other words, in the
contracting phase, the correction coming from inverse powers of the
densitized triad in background dynamics causes vector perturbations to
grow even faster, though only slightly, than in the classical
situation.  The quantum correction depends inversely on volume, i.e.\
it becomes stronger in the smaller volume regime. Using
non-perturbative corrections in $\bar{\alpha}$ for small densitized
triads, a decrease of the rate is indicated since $\bar{\alpha}$ falls
below one in this regime and has $\bar{\alpha}'>0$.  However, for such
small scales the perturbation theory of inhomogeneities is less
reliable and a suppression of the decay rate on very small scales can,
at present, at best be taken as an indication.

\subsection{Quantum corrections: Holonomies}

As before, we consider the situation where anisotropic stress is
absent. Now, the corrected equation for Fourier modes $\sigma^i_k$ of
vector perturbations, after dividing (\ref{QMSigmadotEqnHolo}) by
(\ref{BGDTEoMHolo}), is
\begin{equation} \label{QMGrowthRateHolo}
\frac{{\mathrm d} \log \sigma^i_k}{ {\mathrm d} \log a } = 
- \left(1 + \frac{2\bar{\mu}\gamma\bar{k}}
{\sin 2\bar{\mu}\gamma\bar{k}}\right) ~.
\end{equation}
Here, quantum corrections disappear for $\bar{\mu}\to0$.
Again, the right hand side is less than $-2$ and thus vector modes
grow more strongly in a contracting phase. If the behavior is
extrapolated to the bounce phase, the growth rate in a contracting
universe becomes even larger and would diverge at the bounce where
$\cos(\bar{\mu}\gamma\bar{k})$ is zero. This indicates a breakdown of
the perturbation scheme and the need to include higher order terms as
the bounce is approached. It should also be emphasized here that
$\bar{\mu}\sim 1/{\sqrt{\bar{p}}}$ leads to anomaly free Poisson
algebra only up to order $\bar{k}^4$. So the anomaly term becomes
significant near bounce phase that makes the analysis less reliable
there.

\section{Discussions}

In the absence of anisotropic stress, gauge-invariant vector
perturbations classically grow as $a^{-2}$ in the contracting phase.
Such a growth of vector perturbations indicates a possible violation
of homogeneity assumptions in smaller volume regime, indicated by
the breakdown of classical perturbation theory.  Thus, the growth of
vector perturbations may pose significant problems in particular for
bouncing cosmologies which are invariably associated with a
contracting phase but often have been derived only under the
assumption of homogeneity.  In these models, conclusions regarding
bounces are drawn based on the homogeneity assumption. Naturally, a
growth of vector perturbations can question the robustness of such
bounce scenarios by questioning the validity of the homogeneity
assumption itself at smaller volume, given that bounces are typically
more difficult to realize when inhomogeneities are taken into account;
see \cite{StringInhom,BouncePert}. Thus, it is an important issue to
study the dynamics of vector mode perturbations in the cosmological
context.

In this paper, we have presented a systematic derivation of
gauge-invariant vector perturbation equations
to linear order
in Hamiltonian cosmology
based on Ashtekar variables. 
We have only considered a spatially flat Friedmann--Robertson--Walker
background as this is the case of most interest in cosmology.
Hamiltonians and equations of motion are technically more complicated
in the presence of spatial curvature and are still being worked out.
Quantum corrections are, however, analogous in spatially curved
backgrounds and we do not expect our results to change significantly
in those cases.
Specifically, we have studied the effects of two particular types of
quantum correction, inverse triad and holonomy corrections, on the
dynamics of vector perturbations in large volume regimes. For each
type, we have shown that in a contracting phase the growth rate of
vector mode perturbations is slightly stronger compared to the
classical situation due to quantum effects.  This quantum correction
is small as one expects. Although there are quantization ambiguities,
the sign of corrections to the growth rate seems robust.  A reduction
of the growth rate is indicated only in regimes of non-perturbative
corrections of inverse powers. For such a reduction to be realized,
even if it remains true under a more careful perturbation analysis,
one would have to enter the deep Planck regime.
Moreover, if one starts with a large classical universe as initial
configuration, such non-perturbative quantum effects will become
relevant only after long evolution times. In general, due to the
growth of the vector mode one will eventually have to use higher than
linear orders in perturbative inhomogeneities which we have not
included in this paper. This by itself may well change some of the
conclusions about the bounce phase independently of quantum effects in
the evolution of inhomogeneities.

Another important issue touched in this paper is that of potential
anomalies in the quantum constraint algebra. We started with a general
but unspecified form of a quantum correction function $\alpha$ or
higher order terms, including inhomogeneity.  The functional form of
$\alpha$ as a function of the background variable $\bar{p}$ is known.
However, its functional dependence on the perturbed densitized triad
$\delta E^a_i$ (which is purely off-diagonal for vector modes) is
unknown due to the lack of a systematic derivation of such corrections
from the full theory. As we have observed, requiring anomaly
cancellation in the modified constraint algebra restricts the
functional dependence of quantum correction functions such as $\alpha$
on off-diagonal triad components.  It would be interesting to see
whether such a restriction is satisfied by a systematically derived
quantum correction function from the full theory. As a key result, we
have observed the possibility of non-trivial quantum corrections while
preserving anomaly freedom. The classical constraint algebra for
vector modes is rather trivial, but is much more restrictive for
scalar modes for which the calculations here show the guiding
principle.

\section*{Acknowledgements}

We would like to thank Mikhail Kagan for discussions.
This work was supported in part by NSF grants PHY0653127,
PHY0554771 and PHY0456913.



\end{document}